\documentstyle[preprint,pre,aps,epsfig]{revtex}
\begin{document}
\draft
\title{Dynamics of Small Perturbations of Orbits
on a Torus in a Quasiperiodically Forced 2D Dissipative Map}
\author{Alexey Yu. Jalnine $^{a}$,
Sergey P. Kuznetsov $^{a}$, Andrew H. Osbaldestin $^{b}$
       }
\address{$^{a}$Institute of Radio-Engineering and Electronics of RAS,
Saratov Branch, Zelenaya 38, Saratov 410019, Russia \\
$^{b}$Department of Mathematics, University of Portsmouth,
Portsmouth, PO1 3HE, UK}
\maketitle

\begin{abstract}
We consider the dynamics of small perturbations of stable
two-frequency quasiperiodic orbits on an attracting torus in the
quasiperiodically forced H\'{e}non map. Such dynamics consists in
an exponential decay of the radial component and in a complex
behaviour of the angle component. This behaviour may be two- or
three-frequency quasiperiodicity, or it may be irregular. In the
latter case a graphic image of the dynamics of the perturbation
angle is a fractal object, namely a strange nonchaotic attractor,
which appears in auxiliary map for the angle component. Therefore,
we claim that stable trajectories may approach the attracting
torus either in a regular or in an irregular way. We show that the
transition from quasiperiodic dynamics to chaos in the model
system is preceded by the appearance of an irregular behaviour in
the approach of the perturbed quasiperiodic trajectories to the
smooth attracting torus. We also demonstrate a link between the
evolution operator of the perturbation angle and a
quasiperiodically forced circle mapping of a special form and with
a Harper equation with quasiperiodic potential.
\end{abstract}

\narrowtext
\section{Introduction}
\label{sec:Int}

From the beginning of 1980s, much attention has been paid to
studies of quasiperiodically forced systems. Indeed, systems with
controllable ratios of incommensurate frequencies represent
convenient models for the analysis of bifurcations of
quasiperiodic regimes and of the mechanisms of transition from
quasiperiodicity to chaos. Attention to this class of systems
emerged also due to strange nonchaotic attractors\footnote[1]
{Strange nonchaotic attractors were first described in Ref.
\cite{GOPY}. The term ``strange'' characterizes a fractal-like
geometrical structure of the attractor, while the term
``nonchaotic'' suggests absence of exponential instability of the
trajectories on the attractor. Appearing on the border of
quasiperiodicity and chaos, attractors of this type possess mixed
properties of the both types of dynamics. For more details on
properties of the SNA see Refs.
\cite{DGO,HO,Stark,Kel,KPF,PF1,PZFK,FPP,SW,PF2}.} (SNAs), which
can generically occur there. A large number of works was devoted
to the investigation of mechanisms of regular bifurcations
\cite{Kan,BHTB,CGS,Glen,Daido,FPK,OWGF,Kuz1}, irregular dynamical
transitions \cite{Glen,Daido,FPK,OWGF,Kuz1,NK,DRP,HH,PMR,KLO,YL}
and attractor crises \cite{OF,KL,SFKP,HD,SWS,WFP,PRSS,NPR} in
quasiperiodically forced systems of different nature.

In most cases, interest in the dynamical systems is focused on the
analysis of stationary dynamical regimes and their
transformations. Researchers deliberately leave aside the dynamics
of transient processes. In our humble opinion, such an approach is
not wholly justified, since transient trajectories may visit wide
regions of the phase space before drawing near the attractor. As a
consequence, transient processes may contain extra information
about the structure of invariant sets in the phase space of any
dynamical system. Moreover, a change of the character of the
transient process under variation of the controlling parameters of
the system may precede transformations of the stationary dynamical
regimes. In particular, a change of the character of the transient
process may signal possible bifurcations and crises of attractors
under further variation of the parameters of the
system\footnote[2]{Refer to the node and focus stable fixed points
of a 2D map. Linear analysis of evolution of small perturbations
near such fixed points reveals different character of approach of
transient trajectories to different fixed points. For the stable
fixed points of these two types, a small variation of parameters
may give rise to different local bifurcations: doubling,
pitchfork, inverse saddle-node etc. for nodal fixed points, and
Neimark bifurcation only for focus fixed points.}.

Here, we proceed with the investigation of the bifurcations and
dynamical transitions in quasiperiodically forced systems,
although we focus on dynamics of transient processes and their
transformations. Our main goal is to show that a regular attractor
may have an irregular transient process in its vicinity, and a
complication of the transient may signal imminent destruction of
the regular attractor under variation of the parameters of the
system.

In the present work we consider the dynamics of small
perturbations of stable quasiperiodic orbits with two
incommensurate frequencies on an attracting torus. As a model
system, we consider the quasiperiodically forced H\'{e}non map,
which can be interpreted as a Poincar\'{e} map of a hypothetical
nonlinear oscillator driven by external biharmonic signal with
irrational frequency ratio. A smooth closed invariant curve in the
3D phase space of the model map corresponds to a Poincare section
of a 2D torus in the 4D phase space of such an oscillator.
Therefore, we shall refer to such a smooth closed invariant curve
as a ``torus''. Evolution of a small perturbation of a trajectory
on a stable torus consists of exponentially decaying radial
component of the perturbation (what corresponds to the approach of
the perturbed trajectory to the attracting invariant curve) and
complex behaviour of the angle component of the perturbation,
which characterizes a rotation of the perturbation vector around
the invariant curve. We show that the dynamics of the angle
component of the perturbation may have the character of a two- or
three-frequency quasiperiodicity, or it may be irregular. In the
first two cases a graphic image of such dynamics represents a two-
or three-frequency torus, while in the latter case this image
corresponds to a strange nonchaotic attractor. Note, that the
attractor of the model system remains a smooth torus, while
irregular transient process appears in its vicinity.

We analyse the global structure of the parameter space of the
model system in the region of complex dynamical transitions
consisting in the destruction of the smooth torus accompanied by
the birth of a strange nonchaotic attractor or a divergence of
trajectories. We demonstrate that the appearance of an irregular
(in terms of angle) character of approach of stable trajectories
to the attracting torus plays the role of a ``precursor'' of such
transitions. Namely, the birth of a SNA via torus gradual
fractalization \cite{NK,DRP}, intermittency \cite{PMR,KLO} and
collision with a saddle torus \cite{HH}, as well as the appearance
of trajectory divergence via the collision of an attracting torus
with a fractal basin boundary \cite{OF,KL}, is preceded by the
appearance of irregular dynamics of the perturbation angle. Note
that the attractor of the system remains smooth while perturbed
trajectories in its vicinity already demonstrate ``strange''
behaviour. On the other hand, regular torus bifurcations such as
torus-doubling are accompanied by regular behaviour of the
perturbation angle both before and after the bifurcation. We also
discuss aspects of a link between the perturbation angle dynamics
and a special circle mapping and with the Harper equation with a
quasiperiodic potential.

The paper is organized as follows. In Sec.~\ref{sec:bif} we
briefly observe the types of attractors and dynamical transitions
in the parameter space of the model system. In Sec.~\ref{sec:circ}
we obtain the equation of the evolution of the perturbation angle
in explicit form, and discuss its connection with the Harper
equation (Sec.~\ref{sec:harp}). In Sec.~\ref{sec:dtrans} we
analyse numerically different kinds of the perturbation angle
dynamics and demonstrate their relationship with the types of
dynamical transitions in the parameter space of the model system.

\section{Bifurcations and dynamical transitions in the quasiperiodically forced H\'{e}non map}
\label{sec:bif}

Our model system has the form:
\begin{equation}
\begin{array}{lll}
x_{n+1}=f(x_n,y_n,\theta_n) \equiv a - x_{n}^{2} - b\, y_{n} + \varepsilon \cos{\theta_n}, \\
y_{n+1}=g(x_n,y_n,\theta_n) \equiv x_{n}, \\
\theta_{n+1} = \theta_{n} + 2 \pi \omega \pmod {2\pi},
\end{array}
\label{eq:eq1}
\end{equation}
where the quasiperiodic force frequency parameter is chosen equal
to the inverse golden mean value: $\omega = (\sqrt{5}-1)/2$.

\begin{figure}[!ht]
\begin{center}
\centerline{\epsfig{file=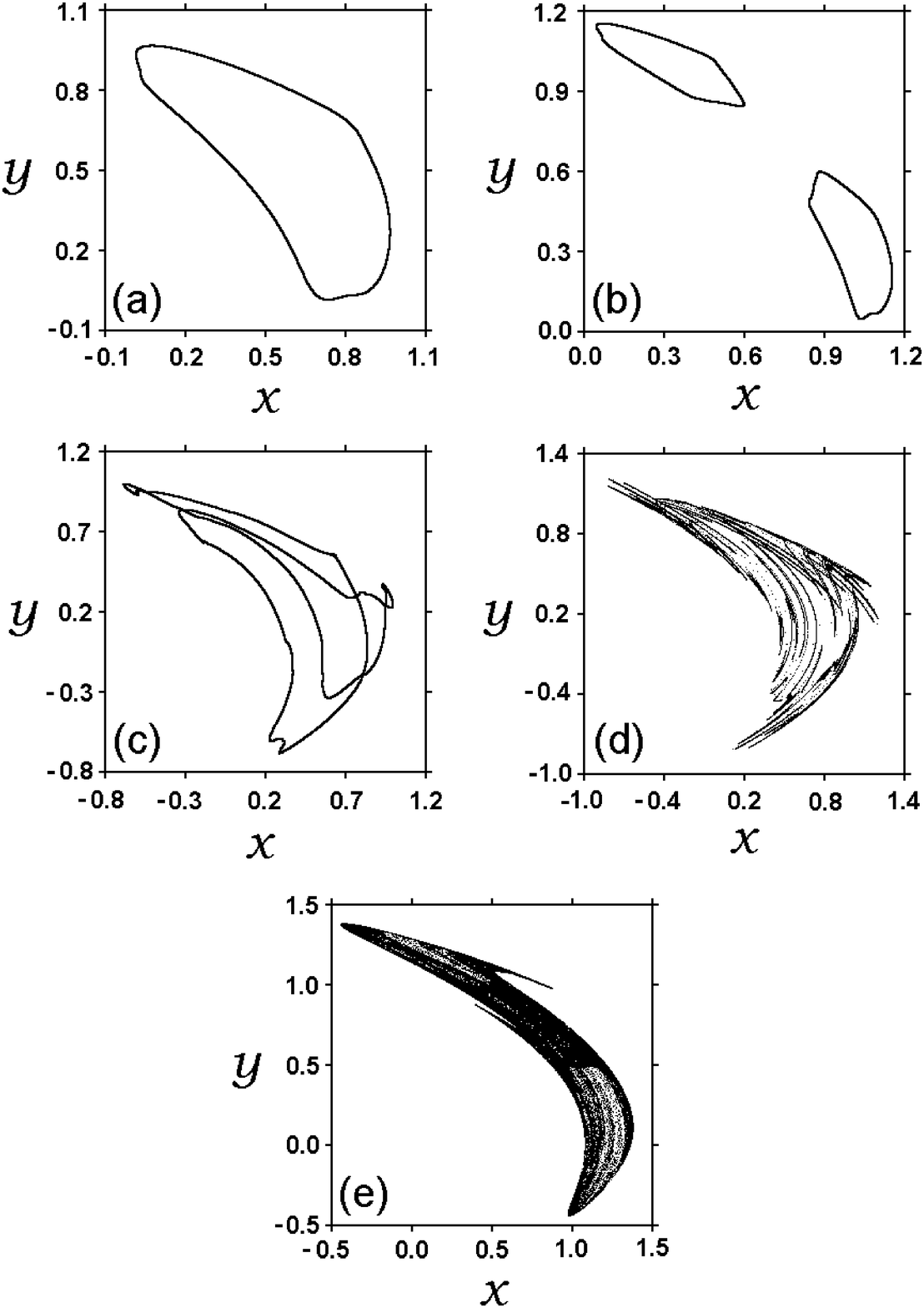, width=0.9\textwidth}}
\caption{The typical attractors of the map~(\ref{eq:eq1}):
(a)~torus T, (b)~double torus 2T, (c)~``double-wrapped'' torus
${\rm T}_2$, (d)~SNA, (e)~chaotic attractor.} \label{fig:f1}
\end{center}
\end{figure}

The examples of typical attractors of the map~(\ref{eq:eq1}) are
presented in Fig.~\ref{fig:f1}. At $a=1.09$, $\varepsilon=0.2$
(here and hereafter we set $b=0.3$) the attractor is a smooth
closed invariant curve (``torus'' T, Fig.~\ref{fig:f1}(a)). With
an increase of the nonlinearity parameter $a$, the torus T may
bifurcate into a ``double torus'' 2T, which consists of a pair of
smooth closed curves mapping into each other under iteration of
the map, as shown in Fig.~\ref{fig:f1}(b) where $a=1.445$,
$\varepsilon=0.095$. (If $\varepsilon$ is small enough, one can
also observe a second torus-doubling bifurcation, which gives four
smooth closed curves. Note, that the number of torus-doubling
bifurcations on the route to chaos depends upon the quasiperiodic
force amplitude $\varepsilon$; this number may increase and tend
to infinity as $\varepsilon$ goes to zero \cite{Kuz2}. The
mechanism for the termination of torus-doubling cascades in
invertible systems is discussed in Ref.~\cite{JO}.) On the other
hand, a different kind of torus-doubling bifurcation (torus
``length-doubling'' bifurcation \cite{HD}) can occur for
sufficiently large values of $\varepsilon$. This bifurcation
produces a special type of torus, which is characterised by the
existence of two wraps of a single invariant curve; we will refer
it to as a ``double-wrapped torus'' ${\rm T}_2$ (see
Fig.~\ref{fig:f1}(c) where $a=0.728$, $\varepsilon=0.36$). Strange
nonchaotic and chaotic attractors of the map~(\ref{eq:eq1}) are
presented in Fig.~\ref{fig:f1}(d) ($a=0.94$, $\varepsilon=0.312$)
and Fig.~\ref{fig:f1}(e) ($a=1.6$, $\varepsilon =0.135$),
respectively.

\begin{figure}[!ht]
\begin{center}
\centerline{\epsfig{file=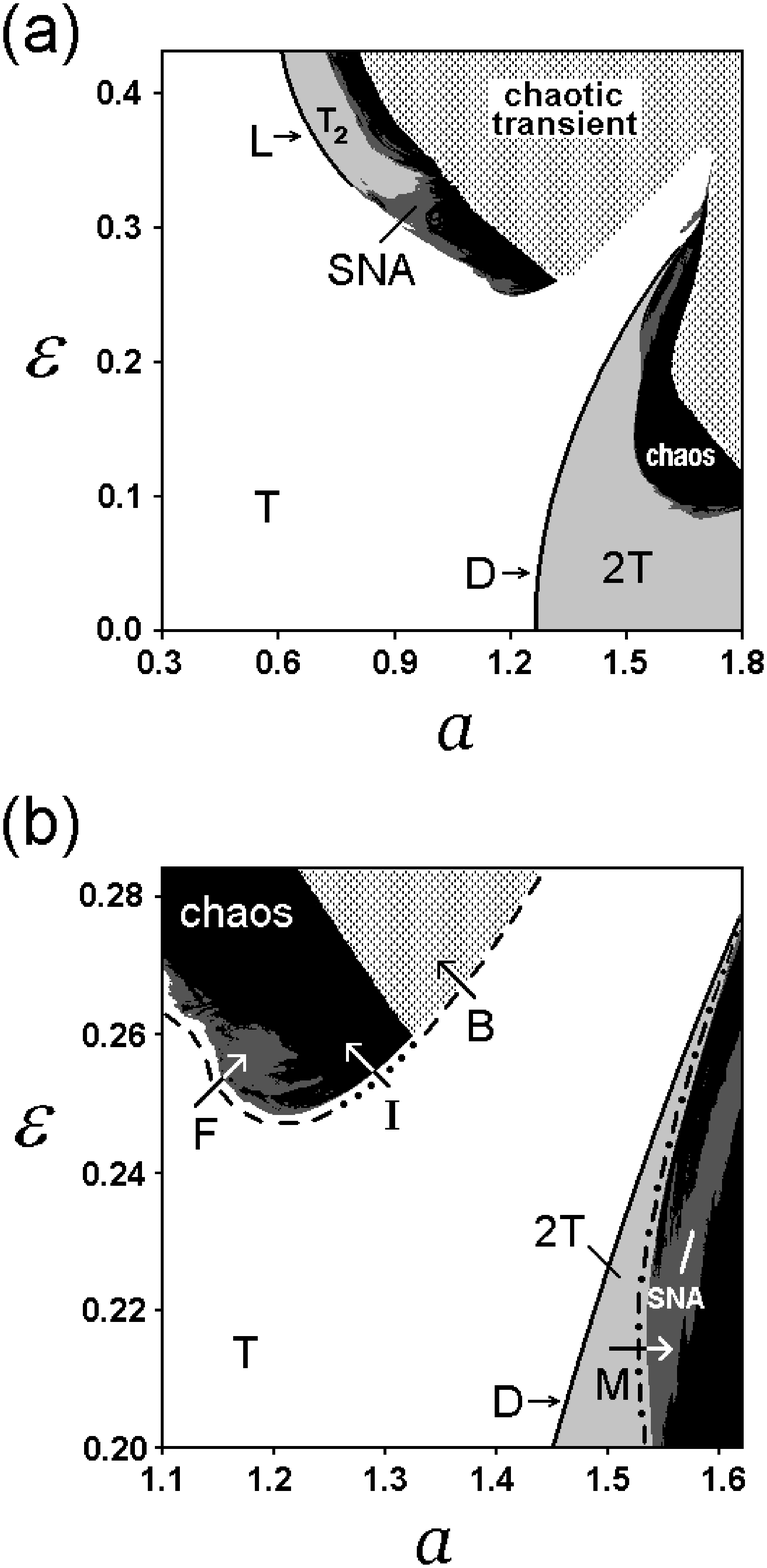, width=0.63\textwidth}}
\caption{(a)~The phase diagram of the map~(\ref{eq:eq1}). (b)~An
enlarged fragment of the same diagram in the region of complex
dynamical transitions.} \label{fig:f2}
\end{center}
\end{figure}

The phase diagram of the map~(\ref{eq:eq1}) in the $a-\varepsilon$
parameter plane ($b=0.3$) is shown in Fig.~\ref{fig:f2}(a). The
regions of different types of dynamical behaviour are shown in
different tones. Torus T (white), double torus 2T (light-grey),
double-wrapped torus ${\rm T}_2$ (light-grey). Regions of chaotic
dynamics are shown in black. On the border between
quasiperiodicity and chaos regions of SNA (dark-grey) exist. In
the patterned area the map~(\ref{eq:eq1}) has no attractor, and
the trajectories go to infinity.

In Fig.~\ref{fig:f2}(b) an enlarged fragment of the phase diagram
in the region of complex dynamical transitions is presented.
Different lines passing along the border of the region of
quasiperiodic dynamics correspond to different dynamical
transitions occurring at the exit from this region. The dashed
line {\bf F} corresponds to the birth of a SNA via gradual
fractalization of the smooth torus T \cite{NK}. The dotted line
{\bf I} corresponds to the intermittent mechanism of the birth of
a SNA \cite{PMR} via collision of torus T with a saddle chaotic
invariant set \cite{KLO}. The combined line {\bf M} corresponds to
the birth of a SNA via band-merging collision of a double torus 2T
with a saddle parent torus \cite{HH}. A transition {\bf B} from
the region of quasiperiodicity to divergence of trajectories
occurs when the torus T destroys and transforms into a chaotic
transient via collision with a saddle chaotic set on the fractal
basin boundary \cite{OF,KL}. Arrows give the direction of movement
in the parameter space for each transition to be observed.

\section{Perturbation angle dynamics and a special circle mapping}
\label{sec:circ}

Let us take a smooth attracting torus of the map~(\ref{eq:eq1}),
given by invariant curve of the form
\begin{equation}
{\rm T}_{ \{a,b\} } :  \{(x,y,\theta) \in {\bf R}^2 \times {\bf
T}^1 | x = X_{\{a,b\}}(\theta), y = Y_{\{a,b\}}(\theta), \theta
\in [0, 2 \pi)\}. \label{eq:eq2}
\end{equation}
The pair of continuous and smooth functions
$\{X_{\{a,b\}}(\theta), Y_{\{a,b\}}(\theta)\}$ can be obtained by
solving the system of nonlinear functional equations
\begin{equation}
\left\{ \begin{array}{ll}
X(\theta + 2 \pi \omega) = a - X^2(\theta) - b\, Y(\theta) + \varepsilon \cos{\theta}, \\
Y(\theta + 2 \pi \omega) = X(\theta). \end{array} \right.
\label{eq:eq3}
\end{equation}
In what follows we can omit mention of the dependence of these
functions upon the parameters $\{a,b\}$ and write simply
$\{X(\theta), Y(\theta)\}$, although we should keep in mind the
link of $\{X(\theta),Y(\theta)\}$ with the system~(\ref{eq:eq3}).

Take a reference quasiperiodic trajectory
$\{(x_n,y_n,\theta_n)\}_{n=0,1,\ldots,\infty} \in {\rm T}$, and
consider a perturbed trajectory close to it starting with  the
same initial phase: $\{(x_n+\delta x_n, y_n+\delta y_n,
\theta_n)\}_{n=0,1,\ldots,\infty}$. Evolution of a small
perturbation $\delta {\bf r}_n = (\delta x_n,\delta y_n)
(n=0,1,\ldots,\infty)$ of the trajectory on torus in linear
approximation is given by the following map
\begin{equation}
\begin {array}{ll}
\left[ \begin{array} {cc} \delta x_{n+1} \\ \delta y_{n+1}
\end{array} \right] = \hat{\bf J}(x_n,y_n,\theta_n) \left[
\begin{array}{cc} \delta x_n \\ \delta y_n \end{array} \right], \\
\theta_{n+1} = \theta_n + 2 \pi \omega \pmod{2 \pi},
\end{array}
\label{eq:eq4}
\end{equation}
where
\begin{equation}
\hat{\bf J}(x,y,\theta) = \left[ \begin{array}{cc}
\frac {\partial f(x,y,\theta)}{\partial x} & \frac{\partial f(x,y,\theta)}{\partial y} \\
\frac{\partial g(x,y,\theta)}{\partial x} & \frac {\partial
g(x,y,\theta)}{\partial y} \end{array} \right]. \label{eq:eq5}
\end{equation}
Coordinates $x_n$ and $y_n$ in the map~(\ref{eq:eq4}) are taken at
the corresponding points of the trajectory on the torus T:
$x_n=X(\theta_n)$, $y_n=Y(\theta_n)$. For convenience, let us
introduce terms
\begin{displaymath}
\begin{array}{ll}
F_x(\theta_n) = \left. \frac{\partial
f(x_n,y_n,\theta_n)}{\partial x_n} \right|_{x_n=X(\theta_n),\;
y_n=Y(\theta_n)}, &F_y(\theta_n) = \left. \frac{\partial
f(x_n,y_n,\theta_n)}{\partial y_n} \right|_{x_n=X(\theta_n),\;
y_n=Y(\theta_n)}, \\
G_x(\theta_n) = \left. \frac{\partial
g(x_n,y_n,\theta_n)}{\partial x_n} \right|_{x_n=X(\theta_n),\;
y_n=Y(\theta_n)}, & G_y(\theta_n) = \left. \frac{\partial
f(x_n,y_n,\theta_n)}{\partial y_n} \right|_{x_n=X(\theta_n),\;
y_n=Y(\theta_n)}.
\end{array}
\end{displaymath}
Now the map~(\ref{eq:eq4}) can be rewritten as
\begin{equation}
\begin{array}{lll}
\delta x_{n+1}=F_x(\theta_n) \delta x_n + F_y(\theta_n) \delta y_n, \\
\delta y_{n+1}=G_x(\theta_n) \delta x_n + G_y(\theta_n) \delta y_n, \\
\theta_{n+1} = \theta_{n} + 2 \pi \omega \pmod {2\pi}.
\end{array}
\label{eq:eq6}
\end{equation}
In polar coordinates
\begin{displaymath}
\delta x = \delta r \cos{\varphi}, \quad \delta y = \delta r
\sin{\varphi}
\end{displaymath}
the map~(\ref{eq:eq6}) becomes
\begin{equation}
\begin{array}{lll}
\delta r_{n+1} \cos{\varphi_{n+1}}=F_x(\theta_n) \delta r_n \cos{\varphi_n} +
F_y(\theta_n) \delta r_n \sin{\varphi_n}, \\
\delta r_{n+1} \sin{\varphi_{n+1}}=G_x(\theta_n) \delta r_n \cos{\varphi_n} +
G_y(\theta_n) \delta r_n \sin{\varphi_n}, \\
\theta_{n+1} = \theta_{n} + 2 \pi \omega \pmod {2\pi}.
\end{array}
\label{eq:eq7}
\end{equation}
Substituting into~(\ref{eq:eq7}) the values
\begin{displaymath}
\begin{array}{ll}
F_x(\theta_n) = -2 X(\theta_n), &F_y(\theta_n) = - b, \\
G_x(\theta_n) = 1, & G_y(\theta_n) = 0,
\end{array}
\end{displaymath}
we have
\begin{equation}
\begin{array}{lll}
\delta r_{n+1} \cos{\varphi_{n+1}}= -\delta r_n [2X(\theta_n)
\cos{\varphi_n} + b \sin{\varphi_n}], \\
\delta r_{n+1} \sin{\varphi_{n+1}}=\delta r_n \cos{\varphi_n}, \\
\theta_{n+1} = \theta_{n} + 2 \pi \omega \pmod {2\pi},
\end{array}
\label{eq:eq8}
\end{equation}
and thus, eliminating the radius variable $\delta r$,  we obtain
\begin{equation}
\begin{array}{ll}
\cot{\varphi_{n+1}}= -[2X(\theta_n) + b \tan{\varphi_n}], \\
\theta_{n+1} = \theta_{n} + 2 \pi \omega \pmod {2\pi}.
\end{array}
\label{eq:eq9}
\end{equation}
In order to get an explicit map for the angle variable $\varphi$,
one should take into account the conditions
\begin{equation}
\begin{array}{ll} \varphi_n \in [0,\pi/2) \cup [3 \pi /2, 2 \pi)
\Rightarrow \varphi_{n+1} \in [0,\pi), \\
\varphi_n \in [\pi /2, 3 \pi /2) \Rightarrow \varphi_{n+1} \in
[\pi, 2 \pi),
\end{array}
\label{eq:eq10}
\end{equation}
which immediately follow from the second equation of the
system~(\ref{eq:eq8}). Then, from~(\ref{eq:eq9})
and~(\ref{eq:eq10}) one can finally obtain the explicit mapping
for the evolution of the perturbation angle variable $\varphi$:
\begin{equation}
\begin{array}{ll}
\varphi_{n+1}= f(\varphi_n,X(\theta_n)), \\
\theta_{n+1} = \theta_{n} + 2 \pi \omega \pmod {2\pi},
\end{array}
\label{eq:eq11}
\end{equation}
where
\begin{displaymath}
f(\varphi,X(\theta)) = \left\{ \begin{array}{ll}
\arctan{[2X(\theta)+b \tan{\varphi}]} + (\pi /2), \quad
\varphi \in [0,\pi/2) \cup [3 \pi /2, 2 \pi), \\
\arctan{[2X(\theta)+b \tan{\varphi}]} + (3 \pi /2), \; \varphi \in
[\pi /2, 3 \pi /2).
\end{array} \right.
\end{displaymath}
Recall that $X(\theta)$ is a function of period $2 \pi$. The
map~(\ref{eq:eq11}) represents a kind of a circle mapping under
external quasiperiodic forcing. Note, that attractors of the
map~(\ref{eq:eq11}) must be symmetric with respect to a shift
$\pi$ along the $\varphi$-axis. Therefore, the map~(\ref{eq:eq11})
could be simplified to the form
\begin{displaymath}
\begin{array}{ll}
\varphi_{n+1}=\arctan{[2X(\theta_n) + b \tan{\varphi_n}]}+(\pi
/2), \\ \theta_{n+1} = \theta_{n} + 2 \pi \omega \pmod {2\pi},
\end{array}
\end{displaymath}
by the choice of $\varphi \in [0, \pi)$.

Returning to the radial coordinate $\delta r$, one can see, that
its dynamics is trivial. Indeed, for any small perturbation in a
vicinity of the attracting torus T we immediately have:
\begin{displaymath}
\delta r_n = \sqrt{(\delta x_n)^2 + (\delta y_n)^2} \sim \exp{(n
\lambda_1 )} \sqrt{(\delta x_0)^2 + (\delta y_0)^2} = \delta r_0
\exp{(n \lambda_1)},
\end{displaymath}
where $\lambda_1<0$ is the largest nontrivial Lyapunov exponent,
which characterizes stability of the torus T.

The analysis of the dynamics of the perturbation angle $\varphi$
on complex tori of the form  $k\,{\rm T} \; (k=2,4,8,\ldots)$ in
the map~(\ref{eq:eq1}) requires trivial modification of the
map~(\ref{eq:eq11}). The last becomes cyclic:
\begin{displaymath}
\begin{array}{ll}
\varphi_{n+1}= f(\varphi_n,X^{(m)}(\theta_n)), \\
\theta_{n+1} = \theta_{n} + 2 \pi \omega \pmod {2\pi},
\end{array}
\end{displaymath}
where $n=kp+m$, $p$ is integer, $m=1,2,\ldots,k$, and
$\{X^{(m)}(\theta)\}$ is a set of functions determining a set of
$k$ smooth closed curves which the complex torus $k\, {\rm T}$
consists of. On the other hand, for analysis of the dynamics of
the perturbation angle $\varphi$ on the double-wrapped torus ${\rm
T}_2$ it is convenient to redefine the phase variable $\theta$ on
the interval $[0,4 \pi)$ and to rewrite the map~(\ref{eq:eq11}) in
the following form:
\begin{displaymath}
\begin{array}{ll}
\varphi_{n+1}= f(\varphi_n,X_{(2)}(\theta_n)), \\
\theta_{n+1} = \theta_{n} + 2 \pi \omega \pmod {4 \pi},
\end{array}
\end{displaymath}
where $X_{(2)}(\theta)$ is a function of the period $4 \pi$
defining the double-wrapped torus ${\rm T}_2$.

\section{A link to the Harper equation}
\label{sec:harp}

Let us make a change of the variable in the map~(\ref{eq:eq9}):
$\tilde{u}_n = \tan{\varphi_n}$. Then, one can write:
\begin{equation}
\begin{array}{ll}
\tilde{u}_{n+1} = - [b \tilde{u}_n + 2 X(\theta_n)]^{-1}, \\
\theta_{n+1} = \theta_{n} + 2 \pi \omega \pmod {2\pi}.
\end{array}
\label{eq:eq12}
\end{equation}
Then, introducing $u_n = b^{1/2} \tilde{u}_n$, one obtains:
\begin{equation}
\begin{array}{ll}
u_{n+1} = - [u_n + 2 b^{-1/2} X(\theta_n)]^{-1}, \\
\theta_{n+1} = \theta_{n} + 2 \pi \omega \pmod {2\pi}.
\end{array}
\label{eq:eq13}
\end{equation}
Since the function $X(\theta)$ is $2 \pi$-periodic, it can be
decomposed into Fourier series:
\begin{equation}
X(\theta) = a_0/2 + \sum_{k=1}^{\infty}{A_k \cos{k(\theta +
\phi_k)}}. \label{eq:eq14}
\end{equation}
Fourier coefficients $\{A_k\}_{k=1 \ldots \infty}$ and
$\{\phi_k\}_{k=1 \ldots \infty}$ can be obtained from numerical
solution of the functional equation~(\ref{eq:eq3}). For the case
of small values of the quasiperiodic force amplitude $\varepsilon$
(see the map~(\ref{eq:eq1})), one may take into account the first
terms of the Fourier series only and obtain
\begin{equation}
\begin{array}{ll}
u_{n+1} = - [u_n + b^{-1/2}a_0 + 2 b^{-1/2} A_1 \cos{(\theta_n + \phi_1)}]^{-1}, \\
\theta_{n+1} = \theta_{n} + 2 \pi \omega \pmod {2\pi}.
\end{array}
\label{eq:eq15}
\end{equation}
Introducing parameters $E = - b^{-1/2} a_0$, $\lambda = b^{-1/2}
A_1$, the map becomes
\begin{equation}
\begin{array}{ll}
u_{n+1} = - [u_n - E + 2 \lambda \cos{(\theta_n + \phi_1)}]^{-1}, \\
\theta_{n+1} = \theta_{n} + 2 \pi \omega \pmod {2\pi},
\end{array}
\label{eq:eq16}
\end{equation}
which reduces to the Harper equation with a quasiperiodic
potential after a standard change of the variables $u_n =
\psi_{n-1} / \psi_n$:
\begin{equation}
\psi_{n+1} + \psi_{n-1} + 2 \lambda \cos{(2 \pi \omega n + \phi)}
\psi_n = E \psi_n,
\label{eq:eq17}
\end{equation}
where $\phi = \theta_0 + \phi_1$ is a phase shift.

The states of the wave function in the Harper equation are known
to be related to the dynamical regimes of the map~(\ref{eq:eq16}).
The extended states in the subcritical region ($\lambda<1$)
correspond to tree-frequency tori, while the localized states in
the supercritical region ($\lambda>1$) are associated with SNAs in
the map~(\ref{eq:eq16}). A two-frequency torus of the
map~(\ref{eq:eq16}) corresponds to the non-normalizable wave
function and to the gap in the energy spectrum.

\begin{figure}[!ht]
\begin{center}
\centerline{\epsfig{file=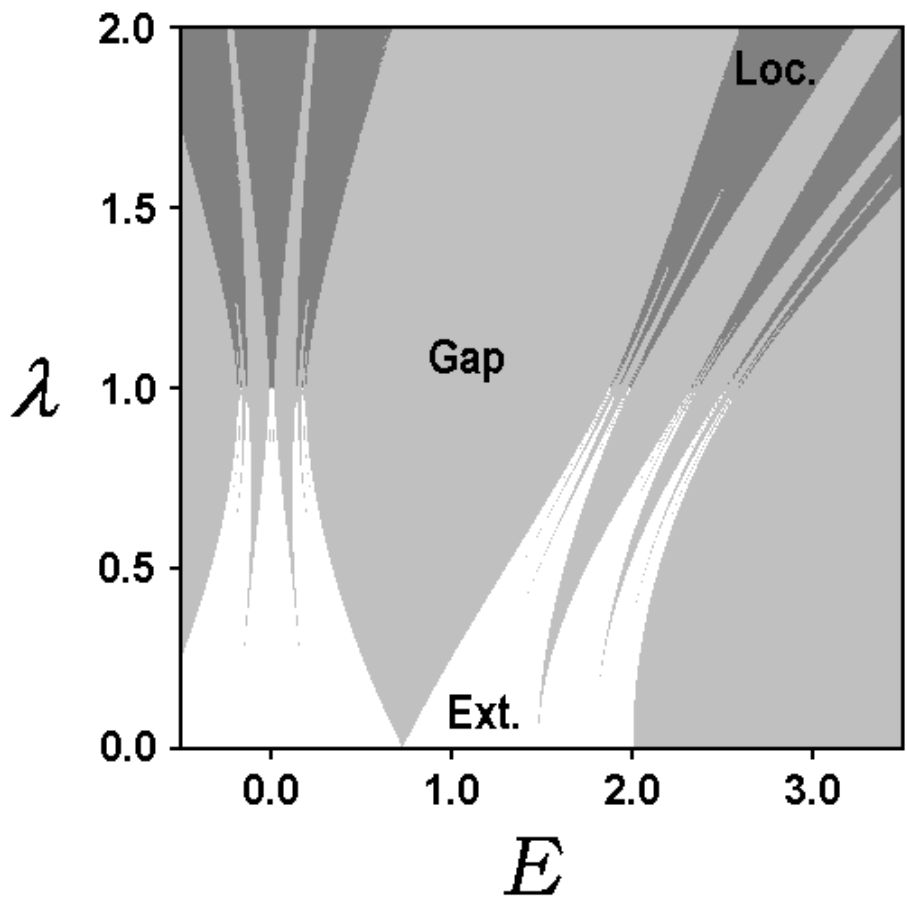, width=0.8\textwidth}}
\caption{The phase diagram of the Harper equation with a
quasiperiodic potential~(\ref{eq:eq17}).} \label{fig:f3}
\end{center}
\end{figure}

In Fig.~\ref{fig:f3} the standard phase diagram of the Harper
map~(\ref{eq:eq16}) is presented. The extended phase (Ext.) is
shown in white, and the localized phase (Loc.) is denoted by gray.
The Gap regions are shown in the light-gray tone. In order to
obtain this diagram, we used the following technique. Each
dynamical regime of the map~(\ref{eq:eq16}) was characterized by
the non-trivial Lyapunov exponent $\sigma$ and the phase
sensitivity exponent $\delta$ \cite{PF2}. These exponents take
values $\sigma = 0$ and $\delta = 0$ for a three-frequency torus,
$\sigma < 0$ and $\delta = 0$ for a two-frequency torus, and
$\sigma < 0$ with $\delta > 0$ for the case of a SNA.

The link from the Harper equation to the map~(\ref{eq:eq11})
suggests that the configuration of the phases on the diagram
(Fig.~\ref{fig:f3}) can be related to the structure of sets in the
parameter space of the quasiperiodically forced H\'{e}non map.
Indeed, in the next section we will show that regions of different
perturbation angle dynamics are organized as ``tongues'' of two-
and three-frequency quasiperiodicity or of a complex behaviour
corresponding to a SNA.

\section{Angular dynamics of perturbations in
the context of torus bifurcations} \label{sec:dtrans}

Now we will focus on the angle dynamics of small perturbations in
a vicinity of the torus T. Suppose we have a trajectory starting
from the initial phase $\theta_0$  on the torus and examine its
small perturbation with an arbitrarily chosen initial angle
$\varphi_0$. Further evolution of the perturbation angle $\varphi$
is determined by the map~(\ref{eq:eq11}). In what follows, we
choose a few sets of the parameters $(a,\varepsilon)$ such that a
smooth torus T exists in the map~(\ref{eq:eq1}) and consider a
dynamics of the map~(\ref{eq:eq11}) at the same parameter values.

\begin{figure}[!ht]
\begin{center}
\centerline{\epsfig{file=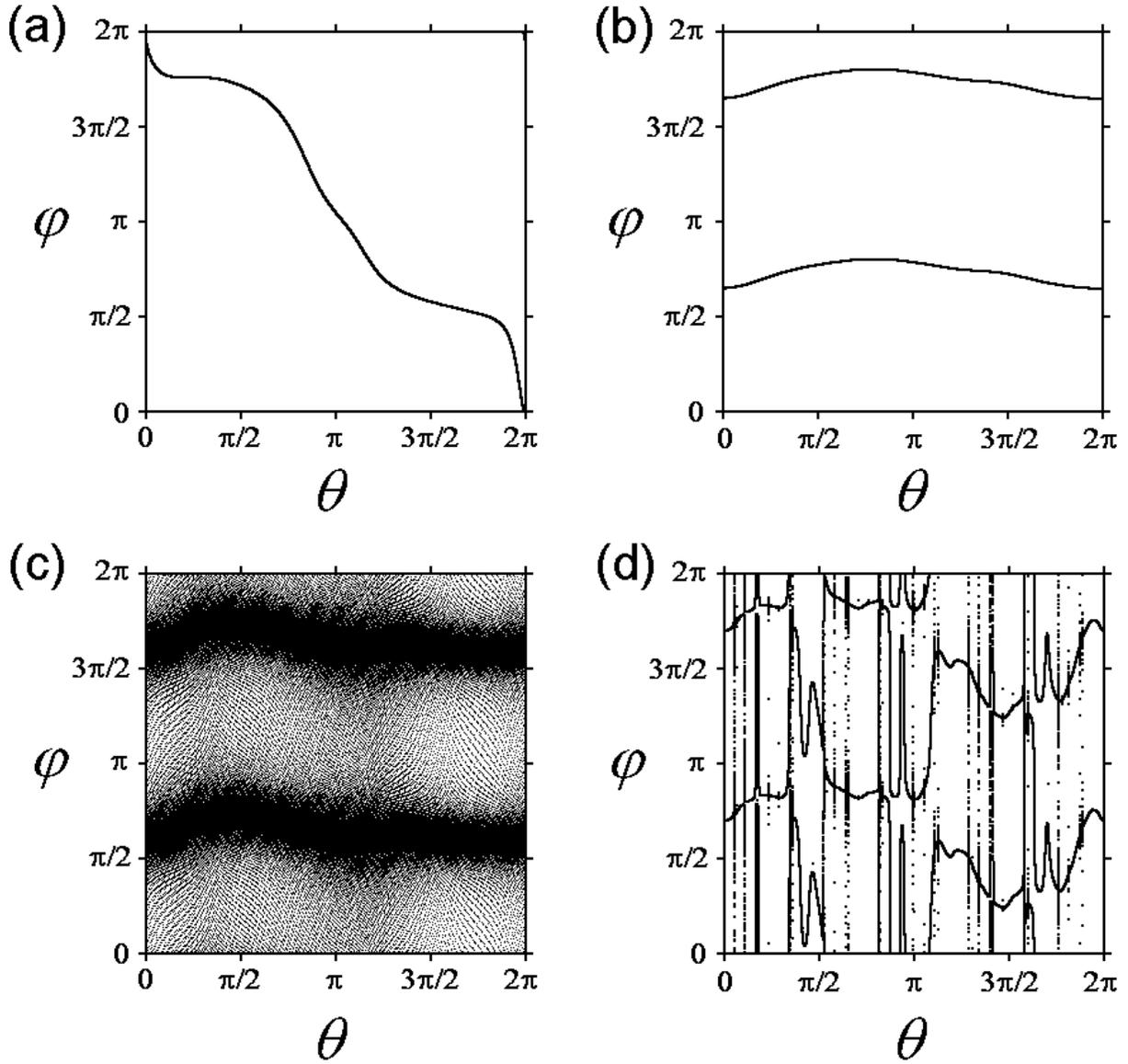, width=\textwidth}}
\caption{Attractors of the map~(\ref{eq:eq11}) as graphic images
of the dynamics of the perturbation angle~$\varphi$: (a),(b)~
two-frequency tori, (c)~tree-frequency torus, (d)~SNA.}
\label{fig:f4}
\end{center}
\end{figure}

At $a=0.8$, $\varepsilon=0.1$ the attractor of the
map~(\ref{eq:eq11}) represents a smooth torus
$\varphi=\Phi(\theta)$ shown in Fig.~\ref{fig:f4}(a). Note that
tori of the map~(\ref{eq:eq11}) may have different topology,
demonstrate one or many wraps on the circle $\varphi \in [0,2
\pi)$ and consist of one or two segments, as shown in the
Fig.~\ref{fig:f4}(b) where $a=1.3$, $\varepsilon=0.5$. The 2D
map~(\ref{eq:eq11}) is characterized by two Lyapunov exponents. A
trivial Lyapunov exponent associated with the quasiperiodic
variable $\theta$  is equal to zero, while a nontrivial Lyapunov
exponent in our cases takes the value $\sigma = -0.1170$ for the
first set of parameter values and $\sigma = -0.9755$  for the
second set. The angle variable $\varphi_n$ of an arbitrarily
chosen perturbation on the torus T asymptotically behaves as
$(\varphi_n - \Phi(\theta_n)) \sim \exp(\sigma n)$, tending to
$\Phi(\theta_n)$ as $n \rightarrow \infty$. Thus, asymptotic
dynamics of the perturbation angle $\varphi_n$ in our cases
represents a regular two-frequency quasiperiodic motion.

As the parameters of the maps are varied to $a=0.85$, $\varepsilon
= 0.077$, the two-frequency torus of the map~(\ref{eq:eq11})
disappears via the backward saddle-node bifurcation, and a
three-frequency torus appears. A trajectory generated by the
map~(\ref{eq:eq11}) then uniformly fills a 2D torus $\{\varphi \in
[0,2 \pi), \theta \in [0,2 \pi)\}$ (see Fig.\ref{fig:f4}(c)).
Strictly speaking, in this situation the map~(\ref{eq:eq11}) does
not possess an attractor, since it is characterized by two zero
Lyapunov exponents. Thus, the dynamics of the perturbation angle
$\varphi_n$ is regular and represents a three-frequency
quasiperiodic motion.

On the other hand, as the parameters change to $a=0.9$,
$\varepsilon=0.28$, the two-frequency torus of the
map~(\ref{eq:eq11}) destroys via phase-dependent mechanism with
the birth of a strange nonchaotic attractor shown in
Fig.~\ref{fig:f4}(d). Note that the nontrivial Lyapunov exponent
remains negative ($\sigma = -0.7747$). The angle variable
$\varphi_n$  of an arbitrarily chosen perturbation in a vicinity
of the torus T asymptotically behaves as $(\varphi_n -
\Phi(\theta_n)) \sim \exp(\sigma n)$ as $n \rightarrow \infty$,
where $\Phi(\theta)$ is a fractal-like function characterized by
non-differentiability and upper/lower semicontinuity \cite{Stark}.
As a result, the dynamics of the perturbation angle $\varphi_n$ is
irregular. A graphic image of such dynamics is a trajectory on a
strange nonchaotic attractor of the map~(\ref{eq:eq11}).

\begin{figure}[!ht]
\begin{center}
\centerline{\epsfig{file=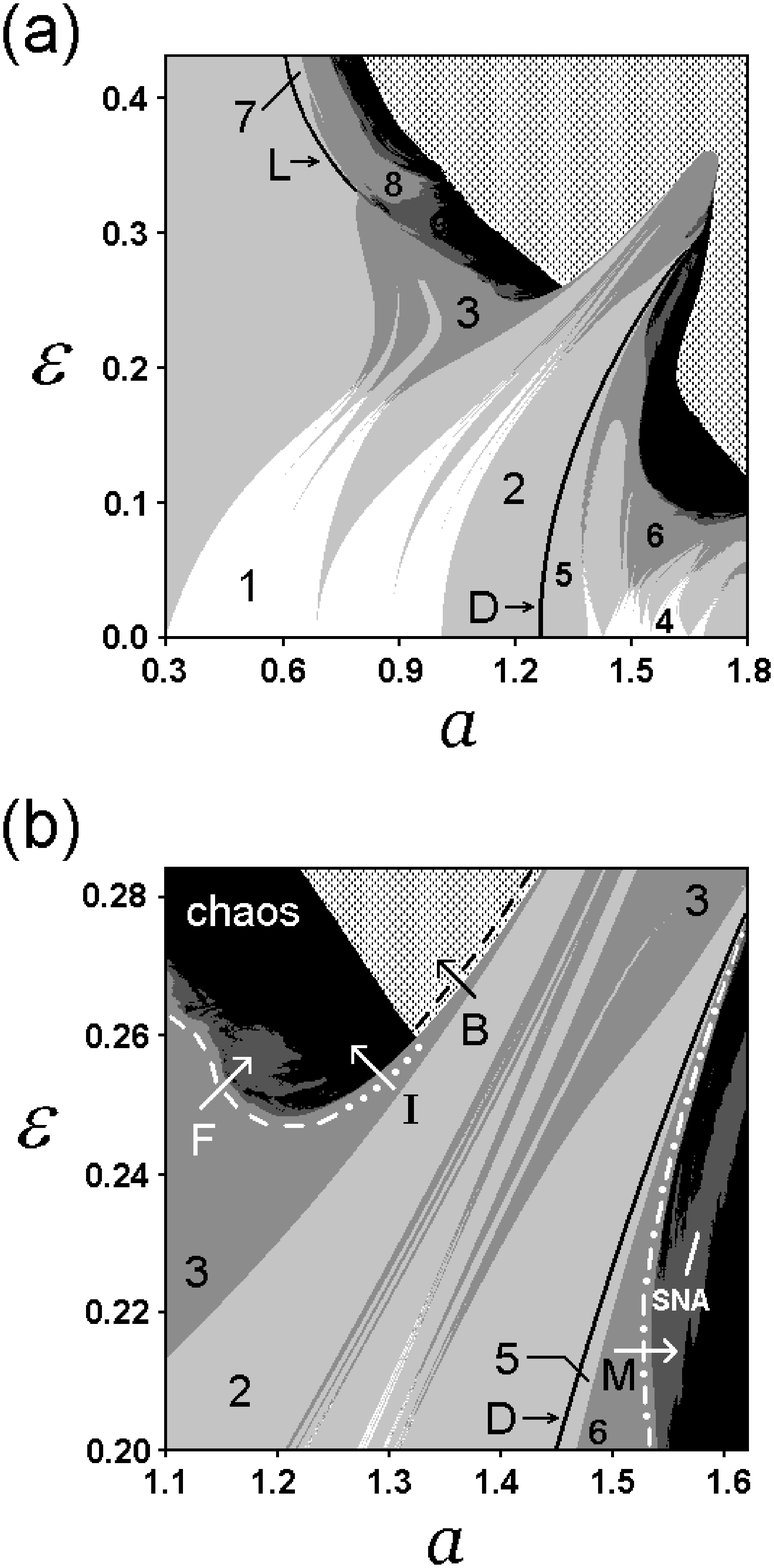, width=0.63\textwidth}}
\caption{The combined phase diagram of the maps~(\ref{eq:eq1}) and
(\ref{eq:eq11}). See text of the Sec.~\ref{sec:dtrans} for
details.} \label{fig:f5}
\end{center}
\end{figure}

In order to observe a configuration of regions corresponding to
different types of dynamics of the perturbation angle variable
$\varphi$ in the $a - \varepsilon$  parameter plane we need to
combine the results obtained for the attractors of the
maps~(\ref{eq:eq1}) and~(\ref{eq:eq11}). Such a combined phase
diagram is presented in Fig.~\ref{fig:f5}(a). The regions of
chaos, SNA and chaotic transient in the map~(\ref{eq:eq1}) are
shown in black, dark-grey and pattern, respectively (the same as
in Fig.~\ref{fig:f2}(a)). The regions of the existence of a smooth
attractor (i.e. T, 2T and ${\rm T}_2$ ) of the map~(\ref{eq:eq1})
are subdivided according to the following principle: the white
tone ({\bf 1},{\bf 4}) denotes regions of three-frequency
quasiperiodic dynamics of the angle variable $\varphi$
(three-frequency torus in the map~(\ref{eq:eq11})), the light-grey
tone ({\bf 2},{\bf 5},{\bf 7}) shows regions of two-frequency
quasiperiodicity of the perturbation angle $\varphi$
(two-frequency torus in the map~(\ref{eq:eq11})), and the grey
tone ({\bf 3},{\bf 6},{\bf 8}) corresponds to regions of irregular
dynamics of the angle $\varphi$  (SNA in the map~(\ref{eq:eq11})).
An enlarged fragment of the diagram in the region of complex
dynamical transitions in the model map~(\ref{eq:eq1}) is shown in
Fig.~\ref{fig:f5}(b).

From Fig.~\ref{fig:f5}(a) one can see that the configuration of
regions of different behaviour of $\varphi$ is rather specific.
The regions of a two- or three-frequency quasiperiodicity and of a
SNA are organized as Arnold tongues. Recall that analogous tongues
of the localized and extended phases were observed for the Harper
equation (Fig.~\ref{fig:f3}). However, the shapes of the tongues
in Fig.~\ref{fig:f5}(a) are distorted compared with those in
Fig.~\ref{fig:f3}. This distortion is associated with the
non-rigorous transition from the Eq.~(\ref{eq:eq13}) to
Eq.~(\ref{eq:eq15}), where only first terms of the Fourier
series~(\ref{eq:eq14}) were taken into account.

The line {\bf D} (${\rm T} \rightarrow 2{\rm T}$) on the phase
diagrams (Fig.~\ref{fig:f5}(a),(b)) correspond to the usual
torus-doubling bifurcation in the map~(\ref{eq:eq1}). The line
{\bf L} (${\rm T} \rightarrow {\rm T}_2$) denotes the torus
length-doubling bifurcation, which leads to the birth of a
double-wrapped torus ${\rm T}_2$.  The lines {\bf F}, {\bf I},
{\bf M} and {\bf B} in Fig.~\ref{fig:f5}(b) denote the same
dynamical transitions on the exit from the quasiperiodicity region
that in Fig.~\ref{fig:f2}(b).

Proceeding from Figs.~\ref{fig:f5}(a),(b), let us analyze the
dynamics of perturbations on the tori T and 2T at the threshold of
the complex transitions {\bf F}, {\bf I}, {\bf M} and {\bf B}. On
the phase diagrams one can see that all the regions of SNA, chaos
and chaotic transient adjoin to the quasiperiodicity regions shown
in the gray tone (and denoted by the numbers {\bf 3} and {\bf 6}),
corresponding to the existence of irregular dynamics of the
perturbation angle variable $\varphi$. Thus, we can claim that the
onset of irregular behavior of the perturbation angle $\varphi$ on
the torus T precedes the following transitions: destruction of a
smooth torus T and the birth of a SNA via fractalization (route
{\bf F}) and intermittency (route {\bf I}) scenarios, destruction
of a smooth torus T and the appearance of a chaotic transient via
collision of the smooth torus T with a fractal basin boundary
(route {\bf B}). In the same way, the appearance of irregular
behavior of the perturbation angle $\varphi$ on the double torus
2T precedes the destruction of the double torus 2T and the birth
of a SNA via collision of 2T with the saddle parent torus (route
{\bf M}). It is important to note that all the transitions
mentioned have phase-dependent mechanisms, as follows from their
analysis in terms of rational approximations \cite{DRP,KLO,KL}.

On the other hand, the lines of regular bifurcations such as torus
doubling ({\bf D}) and torus length-doubling ({\bf L}) separate
regions shown in light-gray tone corresponding to the existence of
two-frequency quasiperiodic dynamics of the perturbation angle
variable $\varphi$. Note, that regular tori bifurcations have
phase-independent character (for instance see Ref.~\cite{BHTB}).
Thus, numerical analysis shows that phase-independent bifurcations
of tori in the model map~(\ref{eq:eq1}) are accompanied by regular
behavior of small perturbations on a torus.

In our opinion, the connection between the behaviour of
perturbations on a torus and the types of torus bifurcations is a
problem of significant mathematical interest. Recently, two of us
(Jalnine and Osbaldestin) suggested a partial solution of this
problem for the case of torus-doubling bifurcation \cite{JO}. We
introduced Lyapunov vectors as basis directions of contraction for
an element of phase volume in a vicinity of the torus and shown
that torus-doubling can occur only if the dependence of the
Lyapunov vectors upon the angle coordinate $\theta$  on torus is
smooth. We have also shown that the appearance of a non-smooth
dependence of the Lyapunov vectors upon the angle coordinate on
torus terminates the line of torus-doubling bifurcation on the
parameter plane. Note that, after sufficiently large number of
iterations, an arbitrarily chosen initial perturbation vector of
the form $\delta {\bf r}_0 = (\delta x_0,\delta y_0, 0)$ on the
torus T (see Sec.~\ref{sec:circ}) in a typical case tends to the
leading Lyapunov vector, i.e., that corresponding to the largest
Lyapunov exponent. Obviously, the regular dynamics of
perturbations on the torus is associated with a smooth dependence
of the Lyapunov vectors upon the angle coordinate $\theta$ on the
torus, while the appearance of a SNA in the dynamics of the
perturbation angle $\varphi$ manifests the onset of a
non-differentiable dependence of the Lyapunov vector upon the
coordinate $\theta$. The latter observation reveals the reason for
regularity of the dynamics of $\varphi$, which accompanies regular
torus bifurcations (e.g. doubling or length-doubling of a torus).
However, the reason for irregularity in the behaviour of $\varphi$
on the threshold of phase-dependent transitions (such as torus
fractalization, intermittency etc.) is still unrevealed and worthy
of further study.

\section{Conclusion}
\label{sec:conc}

In the present work we investigated the dynamics of small
perturbations of trajectories on an attracting torus in the
quasiperiodically forced H\'{e}non map. We have shown that
dynamics of the angle variable of a perturbation may be a two- or
three-frequency quasiperiodicity, or it may have irregular
character. In the last case the graphic image of such dynamics is
a trajectory on a strange nonchaotic attractor. It was also shown
that the appearance of irregular character in the approach of
trajectories to an attracting torus precedes the destruction of
such a torus via phase-dependent mechanisms with the birth of a
strange nonchaotic attractor or of a chaotic transient.

\acknowledgments

The work was supported by the Russian Foundation of Basic Research
(Grant No.~03-02-16074), CRDF (BRHE REC-006 ANNEX BF4M06,
Y2-P-06-16), and by a grant of the UK Royal Society.

\begin{references}
\bibitem{GOPY} C. Grebogi, E. Ott, S. Pelikan, J. A. Yorke,  Physica D {\bf 13}, 261 (1984).
\bibitem{DGO} M. Ding, C. Grebogi, E. Ott, Phys.\ Lett.\ A {\bf 137}, 167 (1989).
\bibitem{HO} B. R. Hunt, E. Ott, Phys.\ Rev.\ Lett.\ {\bf 87}, 254101 (2001).
\bibitem{Stark} J. Stark, Physica D {\bf 109}, 163 (1997).
\bibitem{Kel} G. Keller, Fundamenta\ Mathematicae\ {\bf 151}, 139 (1996).
\bibitem{KPF} S.P. Kuznetsov, A.S. Pikovsky, U. Feudel, Phys.\ Rev.\ E {\bf 51}, R1629 (1995).
\bibitem{PF1} A. Pikovsky, U. Feudel, J.\ Phys.\ A: Math.\ Gen.\ {\bf 27}, 5209 (1994).
\bibitem{PZFK}  A. S. Pikovsky, M. A. Zaks, U. Feudel, J. Kurths, Phys.\ Rev.\ E {\bf 52}, 285 (1995).
\bibitem{FPP} U. Feudel, A. Pikovsky, A. Politi, J.\ Phys.\ A: Math.\ Gen.\ {\bf 29}, 5297 (1996).
\bibitem{SW} J. W. Shuai, K. W. Wong, Phys.\ Rev.\ E {\bf 57}, 5332 (1998).
\bibitem{PF2} A. Pikovsky, U. Feudel, CHAOS {\bf 5}, 253 (1995).
\bibitem{Kan} K. Kaneko, Prog.\ Theor.\ Phys.\ {\bf 69}, 1806 (1983).
\bibitem{BHTB} H. Broer, G. B. Huitema, F. Takens, B. L. J. Braaksma, Mem.\ Amer.\
                Math.\ Soc.\ {\bf 83}, 1 (1990).
\bibitem{CGS} P. R. Chastell, P. A. Glendinning,  J. Stark, Phys.\ Lett.\ A {\bf 200}, 17 (1995).
\bibitem{Glen} P. Glendinning, Discrete\ and\ Continuous\ Dynamical\ Systems\ - Series\ B\,
                Vol. {\bf 6}, No 4, P. 1 (2002).
\bibitem{Daido} H. Daido, Prog.\ Theor.\ Phys.\ {\bf 71}, 402 (1984).
\bibitem{FPK} U. Feudel, A. S. Pikovsky, J. Kurths, Physica\ D {\bf 88}, 176 (1995).
\bibitem{OWGF} H. Osinga, J. Wiersig, P. Glendinning, U. Feudel, Int.\ J.\ of\
Bifurcations\ and\ Chaos\, Vol. {\bf 11}, No 12, P. 3085 (2001).
\bibitem{Kuz1} S. P. Kuznetsov, Phys.\ Rev.\ E {\bf 65}, 066209 (2002).
\bibitem{NK} T. Nishikawa, K. Kaneko, Phys.\ Rev.\ E {\bf 54}, 6114 (1996).
\bibitem{DRP} S. Datta, R. Ramaswamy, A. Prasad, Phys.\ Rev.\ E {\bf 70}, 046203 (2004).
\bibitem{HH} J. F. Heagy, S. M. Hammel, Physica D {\bf 70}, 140 (1994).
\bibitem{PMR} A. Prasad, V. Mehra, R. Ramaswamy, Phys.\ Rev.\ Lett.\ {\bf 79}, 4127 (1997).
\bibitem{KLO} S.-Y. Kim, W. Lim, E. Ott, Phys.\ Rev.\ E {\bf 67}, 056203 (2003).
\bibitem{YL} T. Yalcinkaya, Y.-C. Lai, Phys.\ Rev.\ Lett.\ {\bf 77}, 5039 (1996).
\bibitem{OF} H. M. Osinga, U. Feudel, Physica D {\bf 141}, 54 (2000).
\bibitem{KL} S.-Y. Kim, W. Lim, Phys.\ Lett.\ A {\bf 334}, 160 (2005).
\bibitem{SFKP} O. Sosnovtseva, U. Feudel, J. Kurths, A. Pikovsky, Phys.\ Lett.\ A {\bf 218}, 255 (1996).
\bibitem{HD} J. Heagy, W. L. Ditto, J.\ Nonlinear\ Sci.\ {\bf 1}, 423 (1991).
\bibitem{SWS} J. J. Stagliano, J.-M. Wersmger, E. E. Slaminka, Physica D {\bf 92}, 164 (1996).
\bibitem{WFP} A. Witt, U. Feudel, A. S. Pikovsky, Physica D {\bf 109}, 180 (1997).
\bibitem{PRSS} A. Prasad, R. Ramaswamy, I. I. Satija, N. Shah, Phys.\ Rev.\ Lett.\ {\bf 83}, 4530 (1999).
\bibitem{NPR} S. S. Negi, A. Prasad, R. Ramaswamy, Physica D {\bf 145}, 1 (2000).
\bibitem{Kuz2} S. P. Kuznetsov, JETP\ Lett.\ {\bf 39}, 113 (1984).
\bibitem{JO} A. Yu. Jalnine, A. H. Osbaldestin, Phys.\ Rev.\ E {\bf 71}, 016206 (2005).
\end{references}
\end{document}